# THE MODERATING EFFECT OF GENDER ON ADOPTING DIGITAL GOVERNMENT INNOVATIONS IN ETHIOPIA

Debas Senshaw, Sudan University of Science and Technology & University of Bahir Dar, debassenshaw@gmail.com

Hossana Twinomurinzi, Sudan University of Science and University of Johannesburg, hossanat@uj.ac.za

**Abstracts:** Digital government innovation is being recognised as a solution to many problems faced by governments in providing services to their citizens. It is especially important for low-income countries where there are resource constraints. This research was aimed at exploring the moderating effect of gender on the adoption of a digital government innovation in Ethiopia based on the UTAUT model (n=270) and using structural equation modeling (SEM). The results reveal that gender only moderates the relationship between facilitating conditions and usage behavior of government employees to adopt the digital government innovation which is inconsistent with other findings. Another key finding was that even though the innovation was regarded as not being easy to use, women identified that they would still use it because of the social influence from the peers and the bosses. This finding suggests that women government employees who obtain external support are more likely to use digital government innovations compared with men who are unlikely to use it even if they were facilitated. The paper recommends that governments of low-income countries like Ethiopia should design appropriate policies that encourage women in digital government.

## 1. INTRODUCTION

Digital technology now plays a significant role in governments to offer effective and efficient government services (Flórez-Aristizábal, Cano, Collazos, & Moreira, 2018; Janssen & Estevez, 2013). Digital technology can enhance the quality of government services by reducing the time and other resources that might be lost without the use of it. Digital technology in government can significantly enhance the ease of use as well as the accessibility of services to citizens and organisations to accomplish their goals (Montarnal, Delgado, & Astudillo, 2020; Norris & Reddick, 2013). Digital government has also the power to improve and facilitate the participation of citizens in the political processes and organisations in growing the economy (Chatzoglou, Chatzoudes, & Symeonidis, 2015; Norris & Reddick, 2013).

However, governments in low-income countries struggle to create innovations in digital government despite investing large amounts of capital usually obtained through loans and grant from developed countries and international bodies (Aladwani, 2016; Anthopoulos, Reddick, Giannakidou, & Mavridis, 2016). The development of most government digital innovations such as government web-sites is usually done by digital technology staff without the active participation of citizens or government employees to consider their needs and perceptions on the digital artefact (Abu-Shanab, 2017; Alshehri, Drew, Alhussain, & Alghamdi, 2012).

Another challenge is that many low-income countries do not engage with their local context before creating new digital government artefacts but rather simply adopt digital innovations that had been created elsewhere (Senshaw & Twinomurinzi, 2018). In other words, digital innovations in low-income countries that are based on the local context such as m-pesa (Kim, Zoo, Lee, & Kang, 2017) have been shown to enjoy rapid adoption compared to innovations that have been adopted from other contexts.

The use of local digital government innovations that can provide accurate, useful, and context-based information for government employees is unquestionable especially during hard times of pandemic diseases like the current COVID-19 since they enable to access required information from home (Ahn, Park, Lee, & Hong, 2020).

This study focused on evaluating the influence of gender on adopting locally designed digital government innovations. In this research, gender was considered as a moderator variable due to its significance in technology acceptance models (Suki & Suki, 2017; Venkatesh, Morris, Davis, & Davis, 2003).

Previous research has indicated that men and women decide differently to accept and use digital innovations (Mandari & Chong, 2018; Williams, Roderick, Rana, & Clement, 2016). Gender plays a significant role in predicting usage behavior in information system research (Venkatesh & Davis, 2000; Venkatesh et al., 2003). For instance, the explanatory power of Technology Acceptance Model (TAM) is considerably enhanced after the consideration of gender as a moderating variable (Venkatesh et al., 2003). Particularly, gender has a moderating influence of performance expectancy, effort expectancy, social influence, and facilitating conditions on behavioral intention and use behavior (Venkatesh et al., 2003).

Most of the studies on digital technology adoption are carried out in developed countries (Im, Hong, & Kang, 2011; Liebenberg, Benade, & Ellis, 2018; Meso, Musa, & Mbarika, 2005; Rahman, Jamaludin, & Mahmud, 2011). On the other hand, there are only limited research works relating to the adoption of digital government innovations particularly in the African context.

The study therefore investigated the moderating effect of gender on the acceptance and adoption of digital government innovations. Specifically, the study investigated the moderating influence of gender on the adoption of a digital government innovation created for the WoredaNet in Ethiopia (Raman et al., 2014; Tai & Ku, 2013).

The reason why gender is selected as a moderating factor for this study is because Ethiopia is a patriarchal society along the lines of religion, tradition and culture, with patriarchy embedded in the laws and legislation (Kassa, 2015). This has resulted in disparities between men and women, in the division of labor, the share of benefits, in law and state, in how households are organised, and how these are interrelated (Kassa, 2015). For example, women are underrepresented in leadership and political decision making (Demeke & Gebru, 2015), education (Bayeh, 2016) and in all levels of formal government activities (Alghizzawi, Mahadin, Al-shibly, & Nawafleh, 2020; Naidoo, 2018; Nurhaeni, Yuliarti, & Nugroho, 2016; Özsungur, 2019). The paper, therefore, sought to answer the research question: *How does gender moderate on the adoption of locally designed digital government innovations in the context of resource-constrained low-income countries?*

The remainder of the paper is organised as follows: The next section explains the theoretical background of digital government, the influence of culture on digital government adoption, an overview of the digital government innovation followed by the Unified Theory of Acceptance and Use of Technology (UTAUT). Then, methodology section is presented followed by results and analysis section. Finally, the discussion of results is explained followed by conclusions and limitations of the research.

## 2.　THEORETICAL BACKGROUND

### 2.1.　Digital Government

Many different definitions of digital government are found in different literature based on the context it is used. A digital government is a government digital platform that aims to achieve improved governance (Mettler, 2019). Digital government is also defined as the use of digital technologies that can transform government initiatives and services by enhancing their quality (Misuraca & Pasi, 2019). Digital government plays a significant role in achieving better efficiency of government by changing citizens' behavior (Misuraca & Pasi, 2019; Norris & Reddick, 2013). However, the advantage of digital government can be achieved provided that it is accepted by the users of the digital government (Alshehri et al., 2012). Therefore, investigating the determinants of acceptance and adoption is very important to develop a relevant model.

### 2.2.　The Influence of Culture on Digital Government Adoption

Women in Ethiopia are not treated as men (Bayeh, 2016) and therefore are not benefited from the economic opportunities of the country. Their participation in the economy is therefore limited like many other low-income countries in Africa (Demeke & Gebru, 2015). This is because the orientation of development projects is inclined to men and excludes women from formal employment. This has created an environment where women are engaged in unpaid, tedious household work that forces them to impoverished parts of society (Kassa, 2015). Secondly, they are deprived of access to training, modern technology and access to education though there is certain improvement done by the government of Ethiopia. This study therefore investigated the moderating effect of gender on the adoption of a digital government innovation that was based on the local context.

### 2.3.　Overview of the Adaptive Capability Digital Artefact

The purpose of the app is to provide services particularly for Woredas of Ethiopia that are not using WoredaNet innovatively. Typically, government employees in Woredas are supposed to use the digital innovation to identify opportunities for public service innovations based on the local context. The app was created using design science research (Senshaw & Twinomurinzi, 2020). The need was identified from the low usage of the WoredaNet platform in Ethiopia. The WoredaNet is a digital government platform using fiber and satellite infrastructure across Ethiopia that was implemented by the government of Ethiopia to provide government services to the lowest administrative regions (Woredas). The name WoredaNet comes from "Woreda" which is Amharic. It has the equivalent meaning of a district. Among the services provided by the WoredaNet are video-conferencing, internet, electronic messaging, and voice over IP between federal, regional, and Woreda sites. There are 1,050 Woredas in Ethiopia, 976 of which (93%) have access to the WoredaNet. Despite the access, only a few Woredas actively use the WoredaNet (Miruts & Asfaw, 2014).

The findings of the adaptive capabilities of Woredas that are using WoredaNet innovatively were used to create the app. Adaptive capabilities are capabilities that can introduce a new service or a new way of creating a service (Ogunkoya, 2018; Wilden & Gudergan, 2015). A case study with a qualitative-interpretive approach was adopted to identify the adaptive capabilities of three innovative Woredas (the Amhara Regional State, Dangila town, and Woreta town). The selection of these three Woredas was made by the ICT management of the Amhara Regional State Science, Technology, Information and Communication Commission (STICC). Three public agencies namely, Judiciary (court) office, Human resource office and Finance office were selected from each the three innovative Woredas. The selection criteria were based on their active usage of the WoredaNet government digital platform. To obtain qualitative data, three executives and management (those who oversee similar tasks in government agencies), one from each government agency, one ICT support, and a district administrator (or a representative) from each of the three Woredas were chosen for the interview. A total of 15 interviewees were included. Thematic analysis was used to elicit adaptive capability codes and themes that were used as a requirement to design the digital innovation. Elaborated Action Design Research (EADR) approach (Mullarkey & Hevner, 2018) was followed to create the digital government adaptive capability artefact (Figure. 1).

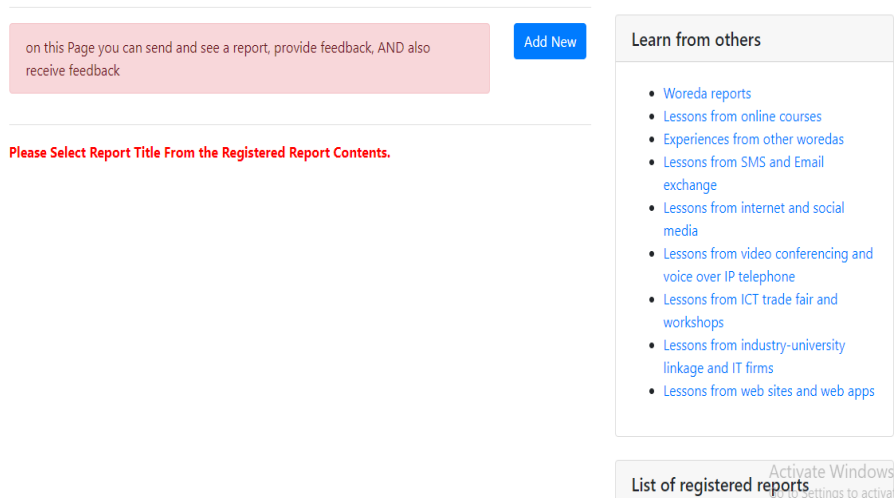

**Figure 1. Digital Government Adaptive Capability Web-based Application**

### 2.4. Unified Theory of Acceptance and Use of Technology (UTAUT)

The Unified Theory of Acceptance and Use of Technology (UTAUT) (Venkatesh et al., 2003) is an enhanced model developed based on the previous technology acceptance models. UTAUT targets to examine the intention of users to use information systems and ultimately enhance their usage behavior through its robust features. UTAUT presents a more comprehensive picture of a technology acceptance process as compared to the previous individual models (Raman et al., 2014; Tai & Ku, 2013).UTAUT integrates the key features of the previous individual models that were used in the field of information systems. UTAUT was selected in this study compared to the newer UTAUT 2 since UTAUT is suited to explicate the attainment of learning tasks in the workplace (Ramírez-Correa, Rondán-Cataluña, Arenas-Gaitán, & Martín-Velicia, 2019). UTAUT is also an important tool for digital government managers to measure the possibility of adoption and acceptance of a new digital government innovation within an organisation, and it enables to examine aspects that help to drive the acceptance of new digital government innovation thereby designing suitable features to enhance the adoption and acceptance of new digital government innovation by the users (Gupta, Dasgupta, & Gupta, 2008).

UTAUT consists of four important constructs, namely, performance expectancy (PE), effort expectancy (EE), social influence (SI), and Facilitating conditions (FC) that are core determinants of behavioral intention (BI) to use technology (Venkatesh et al., 2003). Moreover, UTAUT also uses four variables (experience, gender, age, and voluntariness of use) that can moderate the impact of the four key constructs on behavioral intention and usage of technology.

In this research, a set of hypotheses has been developed and examined regarding the moderating effect of gender on the relationship of the four key constructs and that of the behavioral intention and use behavior of employees to adopt the new digital government innovation.

**Performance Expectancy (PE)**.Performance expectancy refers to the extent to which a user believes that using digital technology enhances his\her job performance (Attuquayefio, 2014). It plays a significant role in validating intention (Agarwal & Prasad, 1998). When users expect digital technology improves their job performance; they tend to accept the technology (Al Khasawneh, 2015; Bhatiasevi, 2016; Chiu, Bool, & Chiu, 2017). Moreover, gender affects the influence of performance expectancy on the intention to use digital technologies (Mandari & Chong, 2018). Therefore, the following hypothesis is tested:

*H1: Performance expectancy positively influences the behavioral intention of government employees to use the app, moderated by gender.*

**Effort Expectancy (EE).** Effort expectancy refers to the extent of ease related to the use of digital technologies (Davis, Bagozzi, & Warshaw, 1989; Lim, Su, & Phang, 2019; Venkatesh et al., 2003). The amount of effort the user puts in affects the behavioral intention of users to accept or adopt digital technologies (Venkatesh et al., 2003). Moreover, gender affects the influence of effort expectancy on the intention to use digital technologies (Venkatesh et al., 2003). Therefore, the following hypothesis is tested:

*H2: Effort expectancy positively influences the behavioral intention of government employees to use the app, moderated by gender.*

**Social Influence (SI).** Social influence refers to the extent to which an individual thinks his/her boss or colleagues believe he or she should use the new digital technology (Venkatesh et al., 2003). Social communication affects users' intention to use digital technologies (Hung, Hsieh, & Huang, 2018). Moreover, gender influences the effect of social influence on behavioral intention to use digital technologies (Venkatesh & Davis, 2000; Venkatesh et al., 2003). Therefore, the following hypothesis is tested:

*H3: Social influence positively influences the behavioral intention of government employees to use the app, moderated by gender.*

**Facilitating Conditions (FC).** Facilitating condition refers to the availability of facilities to use new digital technology (Venkatesh et al., 2003). Unpredictable support of facilities results in lower pressure on the behavioral intention of using digital technologies, whereas predictable support of facilities positively influences use behavior (Attuquayefio, 2014). Moreover, gender influences the effect of facilitating conditions on use behavior (Venkatesh, Thong, & Xu, 2012). This is because men are task-oriented while women are easily affected by the surrounding environment to use digital technologies (Mandari & Chong, 2018). Therefore, the following hypothesis is tested:

*H4: Facilitating condition positively influences government employee's use behavior of the app, moderated by gender.*

## 3. METHODOLOGY

### 3.1. Data Collection and Evaluation

Non-probability sampling technique was implemented to select ten Woredas that have access to the WoredaNet to evaluate the web-based app including the three Woredas that participated in qualitative data analysis. The Amhara Regional State Science, Technology, and Information Communication Commission (STICC) supported the principal researcher mainly for distributing the questionnaire online using Google Forms and deploying the web-based app in the organisation's server at the time of evaluation so that evaluators can easily access the web-based app. It also assisted in assigning the WoredaNet experts in each of the Woredas to facilitate the evaluation response rate. The web-based app was evaluated by government employees consisting of Woreda administrators/representatives, process owners (those who administer similar tasks in public agencies) and experts from ten Woredas' government organisations. The researcher's email and smartphone were used to send the responses.

The questionnaire consisted of a 5-point Likert-scale from strongly disagree to strongly agree and demographic questions. Following the suggestions by (Owoseni & Twinomurinzi, 2018), The twenty items were modified from previous literature to ascertain scale validity as shown in Appendix A. A survey of 400 questionnaires was administered in the beginning. This sample size was decided taking in to account the use of covariance SEM to test the hypotheses with at least 200 sample size (Livote & Wyka, 2009); only 270 complete responses were obtained in 25 days between Dec. 3 to 27, 2019.

The statistical tools SPSS 26 and Amos 26 were used to capture and process the data (Adil, Owais, & Qamar, 2018; Fan et al., 2016; Owoseni & Twinomurinzi, 2018). The data's normality was tested to see whether it met the general linear regression model. The constructs' skewness and kurtosis were determined for each item. The majority of the items were between the acceptable range of -2 and +2 (Abdullah, Chong, Widjaja, & Shahrill, 2017; DeLemos, J. L., Brugge, D., Cajero, M., Downs, M., Durant, J. L., George, Henio-Adeky, S., Nez, T., Manning, T., Rock, T., Seschillie, B., Shuey, & Lewis, 2009), except few items where kurtosis are greater than 2.0 as presented in Appendix B. Overall, the results showed that the skewness and kurtosis values were satisfactory.

From the data the number of women participated in the survey is much fewer than that of the men but adequate to carryout analysis. This was due to a similar sex distribution amongst those who received the questionnaire. Table 1 presents the heterogeneous demographics.

|  | Variable | Frequency | Percent |
|---|---|---|---|
| Gender | Men | 185 | 68.5 |
|  | Women | 85 | 31.5 |
| Experience | 5years and below | 29 | 10.7 |
|  | 6-10years | 123 | 45.6 |
|  | 11-15 | 101 | 37.4 |
|  | 16-20 | 17 | 6.3 |
| Age | 30 and below | 43 | 15.9 |
|  | 31-35 | 121 | 44.8 |
|  | 36-40 | 86 | 31.9 |
|  | 41 and above | 20 | 7.4 |
| Education | Diploma | 35 | 13 |
|  | Bachelor | 186 | 68.9 |
|  | Masters | 49 | 18.1 |

**Table 1. Respondents' Demographic Data**

## 4. RESULT AND ANALYSIS

According to statistical rules, kurtosis and skewness values between -2 and +2 demonstrate a normal univariate distribution (George & Mallery, 2010). Kurtosis is the degree of peackedness of a frequency curve, while skewness is the deviation from symmetry. All skewness values in this analysis are within acceptable limits, and four of the kurtosis values are between +2 and +4 which are satisfactory, indicating the assumption of normality has been met.

For this study, covariance-based structural equation modeling (SEM) using Amos 26 was used to assist with analyzing the data. Covariance-based SEM technique is suitable for exploring models and testing hypotheses based on large sample size (Owoseni & Twinomurinzi, 2018). This is the reason for using it for this study.

The analysis of covariance-based SEM is divided into measurement model analysis and structural model analysis (Owoseni & Twinomurinzi, 2018; Riskinanto, Kelana, & Hilmawan, 2017).The first deals with confirmatory factor analysis (CFA) to examine the reliability and validity of the latent variables while the second would test the hypotheses by examining path coefficients.

### 4.1. Measurement Model

Measurement model analysis focuses on assessment of reliability, convergent validity and discriminant validity (Riskinanto et al., 2017). To assess the model, it is necessary to first scrutinize the value of loading factors of the model's variables if it satisfies the minimum requirement. This was carried out through running Amos 26 software. As shown in Table 2, the constructs' reliability was measured by examining the values of Cronbach's α. The results indicated that the value ranges from 0.745 (for BI) to 0.844 (for SI) which met the minimum requirement of 0.7 (Liao, Fei, & Liu, 2008). The model's reliability was also checked by examining the value of composite reliability. The result indicated that all constructs met the minimum requirement of 0.7 (Rahi, Ngah, & Ghani, 2018). Moreover the reliability of individual items was assessed by investigating the factor loading of all variables with their corresponding constructs. The result indicated that the value of all variables range from 0.522 (for EE4) to 0.870 (for FC3) which met the minimum requirement of 0.5. Thus, the results confirmed that the reliability of the research model was acceptable.

| Construct | Item | Factor loading | P-value | Cronbach's α | CR | AVE |
|---|---|---|---|---|---|---|
| PE: Performance Expectancy | PE1 | 0.660 | *** | 0.836 | 0.844 | 0.578 |
|  | PE2 | 0.839 | *** |  |  |  |
|  | PE3 | 0.847 | *** |  |  |  |
|  | PE4 | 0.674 | *** |  |  |  |
| EE: Effort Expectancy | EE1 | 0.804 | *** | 0.779 | 0.805 | 0.514 |
|  | EE2 | 0.760 | *** |  |  |  |
|  | EE3 | 0.747 | *** |  |  |  |
|  | EE4 | 0.522 | *** |  |  |  |
| SI: Social Influence | SI1 | 0.753 | *** | 0.844 | 0.850 | 0.655 |
|  | SI2 | 0.905 | *** |  |  |  |
|  | SI3 | 0.761 | *** |  |  |  |
| FC: Facilitating Conditions | FC1 | 0.550 | *** | 0.830 | 0.817 | 0.535 |
|  | FC2 | 0.638 | *** |  |  |  |
|  | FC3 | 0.870 | *** |  |  |  |
|  | FC4 | 0.821 | *** |  |  |  |
| BI: Behavioral Intention | BI1 | 0.671 | *** | 0.745 | 0.754 | 0.505 |
|  | BI2 | 0.756 | *** |  |  |  |
|  | BI3 | 0.703 | *** |  |  |  |
| UB: Use behavior | UB1 | 0.783 | *** | 0.807 | 0.810 | 0.681 |
| | UB2 | 0.865 | *** | | | |

Notes:***: Significance at 0.000 Levels, CR=Composite Reliability, AVE= Average Variance Extracted

**Table 2. Reliability and Validity Assessment**

In order to assess the convergent validity, the value of average variance extracted (AVE) was examined for all constructs and met the minimum requirement of 0.5 (Liébana-Cabanillas, Sánchez-Fernández, & Muñoz-Leiva, 2014) as indicated in Table 2 above. The final measurement of the research model was assessing the discriminant validity. This was done by examining the variance shared between the construct and other constructs (Rahi et al., 2018). As illustrated in Table 3, all the square roots of AVE are greater than the corresponding coefficients of correlation with other factors. This shows that all constructs meet the discriminant validity tests. These results indicate that both convergent and discriminant validity analyses have been met. The measurement model ($\chi^2$=234.850, df=154, p-value=0.000), had acceptable fit indices: $\chi^2$/df=1.530, GFI=0.920, TLI=0.956, CFI=0.964, NFI=0.904, RMSEA=0.044.

| Construct | UB | PE | EE | FC | SI | BI |
|---|---|---|---|---|---|---|
| UB | 0.825 | | | | | |
| PE | 0.384 | 0.760 | | | | |
| EE | 0.440 | 0.732 | 0.717 | | | |
| FC | 0.297 | 0.274 | 0.264 | 0.731 | | |
| SI | 0.213 | 0.448 | 0.368 | 0.171 | 0.809 | |
| BI | 0.357 | 0.056 | 0.134 | 0.058 | 0.184 | 0.711 |

**Table 3. Correlation and Square Root of AVEs Matrix**

Due to the acceptable results by CFA, the structural model was developed.

### 4.2. Structural Model

Maximum likelihood, using Amos 26 was applied to measure the structural model. As a result, acceptable fit indices ($\chi^2$/df=1.848, GFI=0.902, TLI=0.928, CFI=0.940, NFI=0.880, RMSEA=0.056) were displayed. Results of regression weights as indicated in the structural model are represented in Table 4. The structural model with chi-square ($\chi^2$) value of 293.786, degree of freedom (df) value of 159, and p-value of 0.000 was displayed. This revealed that the model fitted the data adequately. As a result, it is possible to conclude the research hypotheses using the structural model (shown in Figure 2).

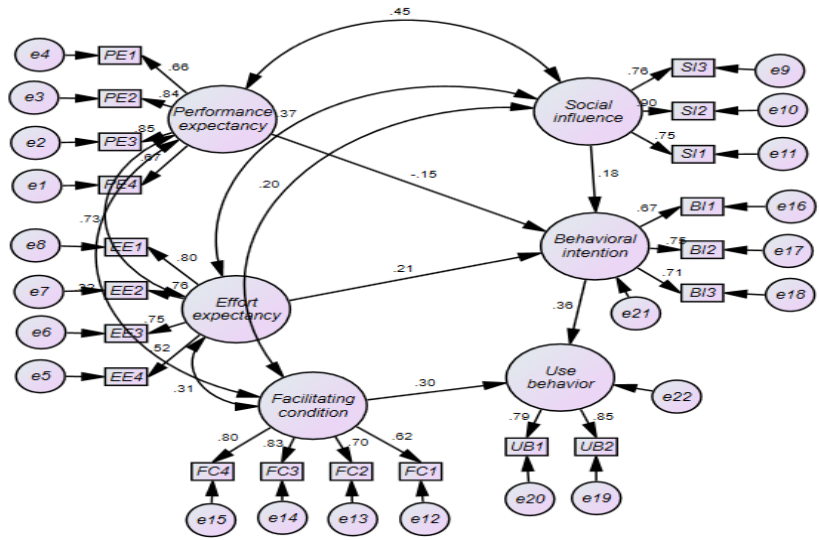

**Figure 2. Structural Model**

The independent and dependent variables will have a significant relationship if the p-value is less than 0.05. It will be insignificant otherwise (see Table 4).

| Independent variable | Relation | Dependent variable | Estimate (β-values) | P-value | Hypothesis conclusion |
|---|---|---|---|---|---|
| Performance expectancy | ⮕ | Behavioral intention | -0.151 | 0.256 | Rejected |
| Social influence | ⮕ | Behavioral intention | 0.184 | 0.033 | Accepted |
| Effort expectancy | ⮕ | Behavioral intention | 0.213 | 0.105 | Rejected |
| Behavioral intention | ⮕ | Use behavior | 0.360 | 0.000 | Accepted |
| Facilitating condition | ⮕ | Use behavior | 0.301 | 0.000 | Accepted |

**Table 4. Regression Weights**

As can be seen from Table 4, the results indicate that performance expectancy ($β=-0.151$, p-value$>0.05$) does not have a significant influence on behavioral intention to accept and use the new digital government innovation. This result is inconsistent with UTAUT (Venkatesh et al., 2003) that examined performance expectancy positively influences behavioral intention to accept and use information systems. The result deviates from the norm as most of the research on technology acceptance is done in developed countries (Im et al., 2011; Rahman et al., 2011). One possible explanation for the inconsistency of the two results is the difference in culture (Al-Gahtani, Hubona, & Wang, 2007; Leidner & Kayworth, 2006), where the two studies are done. UTAUT model was undergone in North America which is culturally different from Ethiopia. Performance expectancy will have less impact on intention specifically on a more collectivism and high power distance culture like Ethiopia where users conform to the expectation of others in higher social roles (Im et al., 2011).

Moreover, the results also reveal that effort expectancy (β=0.213, p-value >0.05) does not have a significant influence on behavioral intention to accept and use the new government digital innovation. This result is also inconsistent with UTAUT (Venkatesh & Davis, 2000; Venkatesh et al., 2003) that examined effort expectancy is a significant predictor of intention. The possible explanation for this inconsistency is the difference in culture where the two studies are carried out (Al-Gahtani et al., 2007; Leidner & Kayworth, 2006). Similarly, it can be explained that the influence of effort expectancy will be weaker in a more collectivism and high power distance culture like Ethiopia where users decision to accept and use new technology is significantly affected by other factors such as social norms (Im et al., 2011).

On the other hand, social influence (β=0.184, p-value <0.05) has a positive impact on the intention to accept and use the new digital government innovation. This means that when employees are encouraged by their boss or colleagues, users will tend to use the new digital government innovation. This is also consistent with UTAUT (Venkatesh et al., 2003) that examined social influence positively influences intention.

The results also show that behavioral intention (β=0.360, p-value <0.05) has a positive influence on the usage of the new digital government innovation. This means that when employees have more intention to use digital innovation, they regularly use it. This result is consistent with UTAUT (Venkatesh et al., 2003) that examined behavioral intention has a significant influence on the usage of information systems.

The results show that facilitating conditions (β=0.301, p-value <0.05) positively influences the usage of the new digital government innovation. When employees are supported with appropriate resources to use digital innovation, they frequently use it. The result is consistent with UTAUT (Venkatesh et al., 2003) that explored facilitating conditions positively influences the usage of information systems.

### 4.3. Moderating Effect

Maximum likelihood using Amos 26 was used to test the influence of gender on the relationship between performance expectancy and behavioral intention, effort expectancy and behavioral intention, social influence and behavioral intention, and facilitating condition and use behavior of government employees to accept and use the digital government innovation. The moderating variable helps to demonstrate if it could affect the direction of the relation or the strength of the relationship between the dependent and independent variables (Hung et al., 2018).To evaluate the moderating effect, cross-multiplication of the observed variables, and moderating variable, gender was carried out. As a result, the standard regression coefficients (β-values) were obtained as indicated in Table 5 below. The results reveal that the regression coefficients were insignificant except that of the interaction of gender and facilitating conditions (β=0.160, p=0.026).

The result reveals that gender is a moderating variable for the relationship between facilitating condition and use behavior. External support to use the digital innovation makes a significant difference among men and women government employees. On the other hand, gender does not have a significance influence on the relationship between performance expectancy and behavioral intention (β=-0.002, p=0.976). This means that there is no difference in the behavioral intention of men and women government employees towards the usefulness of the new digital innovation. The result is inconsistent with UTAUT (Venkatesh et al., 2003). This inconsistency could be created due to the influence of culture as indicated by Baker, Al-Gahtani & Hubona ( 2007). Also, gender does not have influence on the relationship of effort expectancy and behavioral intention (β=0.035, p=0.647). This shows that there is no significant difference among men and women government employees on the ease of use of the web app. This contradicts with the result of UTAUT (Venkatesh et al., 2003) that explored gender as a moderating factor in the relationship between effort expectancy and behavioral intention. The explanation may be men and women in this study have almost similar educational background and work experience that could not make a difference in using the web app. Moreover, gender does not have influence on the relationship between social influence and behavioral intention (β=0.077, p=0.289). This means that social influence does not create a significant difference in the behavioral intention among men and women government employees to use the web app. This is inconsistent with the result of UTAUT (Venkatesh et al., 2003). One explanation may be in a more collectivistic culture like Ethiopia, technology adoption is highly affected by the social influences irrespective of gender (Faqih & Jaradat, 2015).

| Hypothesis | Relationship | Std. Beta | p-value | Decision |
|---|---|---|---|---|
| H1 | PE x Gender→BI | -0.002 | 0.976 | Rejected |
| H2 | EE x Gender→BI | 0.035 | 0.647 | Rejected |
| H3 | SI x Gender→BI | 0.077 | 0.289 | Rejected |
| H4 | FC x Gender→UB | 0.160 | 0.026* | Accepted |

*: Significance at 0.05 levels

**Table 5: Moderator Analysis Results**

This research further explored which group of gender (men or women) produced the highest impact. A multi-group analysis based on gender was implemented to examine the influence of each group on the moderation effect. As a result, the standardized regression coefficients (β-values) of the two groups are presented in Table 6 below. The result implies that the men group has a dampening effect (β=-0.199, P=0.016) on the relationship between facilitating conditions and use behavior since β-value is negative. While the positive moderation effect of the women group was found to be significant on the relationship between facilitating conditions and use behavior with positive β-value (β=0.429, p=0.025). The interpretation is indicated in the discussion section below.

| Gender | Relationship | β -value | p-value | Impact |
|---|---|---|---|---|
| Men | FC→UB | -0.199 | 0.016* | Dampening effect |
| Women | FC→UB | 0.429 | 0.025* | Positive moderating effect |

*: Significance at 0.05 levels

**Table 6: The Effect of Gender on the Moderation Effect**

## 5. DISCUSSION OF RESULTS

The impact of government employees' gender on the four main constructs of the UTAUT model for accepting and using government web app was investigated.

Overall, the digital government innovation was regarded as not being easy to use, mainly because no training was offered. It was therefore understandable that the users did not believe the web app could assist them to perform their jobs better. Nonetheless, they identified that they would still use it because of the social influence from the peers and the bosses.

The results show that gender has a moderating effect on the relationship between facilitating conditions and use behavior of government employees to use and accept the new digital government innovation. The women group was found to have a positive moderation effect on the relationship between facilitating conditions and use behavior with positive β-value (β =0.429, p=0.025). This means that when women government employees obtain more external support to use digital innovation, they are more likely to adopt it compared with men who were unlikely to adopt the digital innovation even if they were facilitated, and would actually use it less.

The result is consistent with the finding of Faqih & Jaradat (2015). One explanation presented is the psychological differences among women and men where external supportive factors are more important to women than they are to men.

The results of this study imply that governments of low-income countries such as Ethiopia should design appropriate policy issues that can encourage women in using local digital government innovations.

The paper contributes to practice and policy in revealing how governments of low-income countries should consider external supports and facilities while implementing locally designed digital government innovations.

## 6. CONCLUSION

The purpose of the study was to explore the moderating effect of gender on the acceptance and use of a digital government innovation. UTAUT model was used to examine the influence of gender on the relationship between the independent and dependent constructs. Moreover, structural equation modeling (SEM) using Amos 26 was adopted to examine the moderating effect of gender. The results reveal that gender has a moderating effect on the relationship between facilitating condition and use behavior of government employees to accept and use the government web-based app, while it does not affect the other relationships. Multi-group analysis was used to examine the moderating effect of gender on the relationship between facilitating condition and use behavior of government employees that use the digital government innovation. The result shows that the women group has a positive influence on moderating the relationship between facilitating condition and use behavior.

### 6.1. Limitation and Future Research

The research was limited to using only gender as a moderating factor on a small sample of Woredas in the Amhara Regional States of Ethiopia. Future studies should also consider increasing the sample Woredas from different regional states and different stakeholders across Woredas.

# Appendix A

| Construct | | Question items |
|---|---|---|
| PE | PE1 | I find the web-based app useful to learn about what others are doing. |
| | PE2 | Using the web-based app increases my chances of getting important information. |
|    | PE3 | Using the web-based app helps me obtain important information more quickly. |
|    | PE4 | Using the web-based app increases my productivity at work.                 |
| EE | EE1 | Learning how to use the web-based app is easy for me.                      |
|    | EE2 | My interaction with the web-based app is clear and understandable.         |
|    | EE3 | I find the web-based app easy to use.                                      |
|    | EE4 | It is easy for me to become skillful at using the web-based app.           |
| SI | SI1 | People who are important to me think that I should use the web-based app.  |
|    | SI2 | People who influence my behavior think that I should use the web-based app. |
|    | SI3 | People whose opinions that I value prefer that I use the web-based app.    |
| FC | FC1 | I have the resources necessary to use the web-based app.                   |
|    | FC2 | I have the knowledge necessary to use the web-based app.                   |
|    | FC3 | The web-based app is compatible with other technologies I use.             |
|    | FC4 | I can get help from others when I have difficulties using the web-based app. |
| BI | BI1 | I intend to continue using the web-based app in the future.                |
|    | BI2 | I will always try to use the web-based app in my daily life.               |
|    | BI3 | I plan to continue to use the web-based app frequently.                    |
| UB | UB1 | I have ever used a web-based app.                                          |
|    | UB2 | I often use web-based apps.                                                |

Note: PE=performance expectancy, EE=effort expectancy, SI=social influence, FC=facilitating condition, BI=behavioral intention, UB=use behavior

**Table 7. Question Items of the Survey Adapted from** (Tan, 2013)

# Appendix B

| Variables (constructs)  | Item | Skew   | Kurtosis |
|-------------------------|------|--------|----------|
| Performance Expectancy  | PE1  | -1.190 | 3.944*   |
|                         | PE2  | -0.909 | 2.381*   |
|                         | PE3  | -0.631 | 0.704    |
| | PE4 | -0.672 | 1.261 |
| Effort Expectancy | EE1 | -0.326 | 0.013 |
| | EE2 | -0.372 | -0.007 |
| | EE3 | -0.756 | 2.732* |
| | EE4 | -0.930 | 2.213* |
| Social Influence | SI1 | -0.932 | 0.849 |
| | SI2 | -0.811 | 0.372 |
| | SI3 | -0.862 | 0.362 |
| Facilitating Conditions | FC1 | -0.494 | 0.008 |
| | FC2 | -0.385 | -0.036 |
| | FC3 | -0.740 | 0.840 |
| | FC4 | -0.603 | 0.056 |
| Behavioral Intention | BI1 | -0.555 | 0.123 |
| | BI2 | -0.807 | 0.925 |
| | BI3 | -0.730 | 0.512 |
| Use Behavior | UB1 | -0.654 | 0.741 |
| | UB2 | -0.647 | 0.751 |

* Kurtosis out of -2 to +2 range

**Table 8. Normality Assessment**